**Short Paper**

# Architectural Visualization Using Virtual Reality: A User Experience in Simulating Buildings of a Community College in Bukidnon, Philippines

Benzar Glen S. Grepon
Program Head, Information Technology Department, Northern Bukidnon State College, Philippines
ben.it2c@gmail.com
(corresponding author)

Aldwin Lester M. Martinez
Information Technology Department, Northern Bukidnon State College, Philippines

**Abstract**

*Purpose* – The study aims to design and develop a virtual structural design that simulates the campus and its buildings of a community college in Bukidnon, Philippines through Virtual Reality. With the immersion of technology, this project represents the architectural design of the establishment with the use of Virtual Reality Technology.

*Method* – The project uses a modified Iterative Development Model which is a guide for the design and development of the 3D Models and VR Application. TinkerCAD which is a web-based application has been used to design buildings on the other hand Unity is used to develop the structural designs of the buildings.

*Results* – The respondents of this study are the Grade 12 Senior High students from the 4 schools which are geographically near to the college. With this study, the researchers were able to showcase its VR Application to the students and later evaluated using a System Usability Scale, a 10 item questionnaire measuring usability with an overall average of 90% or Point Score of 4.5 which is interpreted as excellent in a Likert table for descriptive interpretation. With the use of the VR application potential students of the college will be able to visualize and experience the present structures of the college without being physically present in the area.

*Conclusion* – In this paper, the buildings and structures of NBCC were designed and developed through a Virtual Reality Platform allowing students from different secondary schools that are geographically near to the college to experience the feeling to be in the school without being able to set a step in physically. Using VR Gadgets in navigating buildings is still new from the community which makes the VR application a hit to those who use it. The Application was evaluated personally by every student from different schools and rated excellent. With this advancement of technologies, VR plays a vital part in allowing people to see what's inside the building and navigate around it without being physically present in the place.

*Recommendations* – This study is limited only to simulating the architectural buildings of a community college in northern Bukidnon Philippines and currently has no controllers used and no sound effects. The researchers recommend continuing to improve and developing the project since the school buildings expand every year, adding details to every room, including furniture, add doors to the rooms, landscapes, mini-map UI and improve the landform to be more realistic.

*Research Implications* – The study enabled viewing and designing of school structures using the technology of Virtual Reality. With this study, future development of the college in terms of structural design will be easier to visualize.

## INTRODUCTION

Technology has a lot to do especially in changing our ways of living (Langridge, 2020). New technologies unfold every day, one technology is Virtual Reality (VR) which is a prime player in this generation especially the millennials (Voxburner, 2017). Some applications of Virtual Reality are Gaming, Education, or Architectural Simulation (Castronovo et al., 2019). The development of VR Technologies transitions from desktop-based to mobile ones with additional features that include enhanced involvement and interaction abilities (Wang et al., 2018). People tend to use the conventional method of visualizing objects using Virtual Reality, however, experiencing Virtual Reality in terms of using body motions is another exciting idea to look forward to. Virtual reality has recently become an affordable technology and will contribute to scientific discovery (El Beheiry, et al., 2019). Schools also believe that Virtual Reality as immersed in education and use it as an alternative way of teaching and learning allows students to work out their creativity skills effectively, especially if they are also engaged to design architectural plans and objects, VR plays an important role in teaching process because it provides an interesting and engaging way of acquiring information (Kamińska et al., 2019).

In Japan, a virtual game has been developed from a light novel written by Reki Kawahara entitled *Sword Art Online*. The game will not be played with a traditional controller, but by the movement of any body parts, making the whole project different from any other, with nothing similar available on the market as of now. *Sword Art Online: The Beginning* will come with some additional features that sound more interesting on paper such as systems called Cognitive System and *SoftLayer*, which promise to create an environment that will feel like the real world. Additionally, a proposal of using an integrated adoption model of VR games in order to provide a better understanding of player's perspectives toward VR games which are advancement and improvement in the Electronic Game World (Jang & Park, 2019).

Virtual Reality (VR) is the use of computer technology to create a simulated environment unlike traditional user interfaces, VR places the user inside an experience. Instead of viewing a screen in front of them, users are immersed and able to interact with 3D worlds. By simulating as many senses as possible, such as vision, hearing, touching, even smelling, the computer is transformed into a gatekeeper to this artificial world the only limits to near-real VR experiences are the availability of content and cheap computing power (Jackson, 2018).

Northern Bukidnon Community College is on its way to developing its establishments, particularly on developing its buildings. However, the project focuses only on the enhancement of the existing structures on the campus. The college provides only a miniature model of the campus development plan which was a futuristic plan and not the recent vicinity of the college with the use of blocks, paper, woods, and other variety of materials to create a physical type of architectural model to show the over-all vicinity area of the campus, but it is limited on viewing the external structures of each building on a top view perspective. During school promotion conducted by the college to nearby secondary schools, the admissions office has difficulties in describing the school campus and its buildings especially when a curious student inquires which building and what particular place to go when they are on the campus.

The project came up with an idea on using Virtual Reality (VR) Technology in viewing the structure of an architectural design not only the external but also the internal structure with the near accurate scale allowing participants to move around in the virtual world and can see it from different angles, reach into it, grab it and reshape it. This study aims to develop an android application that will serve as an alternative way on promoting the college and its campus through Architectural Visualization using Virtual Reality to tour prospective students remotely and allows users to view the present and actual architectural structure of the school buildings without being physically present in the area, which can be navigated using a smartphone that runs an android operating system and any VR headset. This technology helps students to be more familiar with the school campus and buildings before they will take entrance examinations and physically inquire about admission requirements. This VR Architectural Application is evaluated through a System Usability Scale (SUS) to measure sustainability and assess functionality for the application to be used during an annual school to school promotion and career guidance for prospective college freshmen of the college.

## LITERATURE REVIEW

Virtual reality has been used in games, education, and architectural simulations. Imagine roaming around a virtual building or a house or a condo development even before construction begins. Looking to the 21st-century learners, teaching and learning are best delivered through virtual reality especially integrating and utilizing it for the educational system here in the Philippines. A Department of Trade and Industry (DTI) sponsored event showcases different ways to engage the student and help them to compete and to align skills and learnings to different schools to near countries as the main focal point of the event, there are sharing of possible opportunities of using VR as a tool for better learning and boost students' interest towards technology (Barrozo et al., 2017). According to Espiritu (2017), the Philippines is very much behind in education, unlike the other countries. Filipino students exert a lot of effort and spend a lot of time doing school-related activities but very little time in laboratories. In virtual reality, common problems such as corrosions and being burn are not an issue since they are not physically involved with what is happening to the virtual environment. Virtual Reality became popular in creating an architectural simulation for business and it can be used also for educational purposes. By this, it gives the researchers the purpose to pursue their project to develop an architectural simulation through virtual reality for Northern Bukidnon Community College.

Japan has created ways to deliver made-to-order designs that are unique to businesses related to housing. Merging Virtual reality into architect's showrooms allows a client to interact and at the same time collaborate ideas on how to improve the current designs and to view the possible visualization of the plan set by the professionals on their traditional blueprints and models. There is a huge demand from them to show possible homes designs through VR Technology since they found that they will see a more realistic view of these designs through VR Technology, one exciting part of using it is they can "walk around a home" even if it's still on the initial stages of the design process, with it, clients have already ideas on the visions set by the architect and also to allow them to make specific arrangements on request they would like to include in their future homes (Kusano, 2017).

Virtual reality has improved and transformed the art of architecture, allowing better insights in terms of "virtual property inspections" since it provides a clearer vision or view of a perspective of a home or property both from the owner and the crews before the construction begins. Using VR Gadgets including the Headset and other materials gives the viewer the ability to fully see a unique experience of what would be the actual property would look like, the most accurate illustration of a traditional blueprint of the proposed project which is an epitome to most architects who would like to have engagement with their clients for more client-centered designs (Rodriguez, 2018).

In line with the new development of the Internet of Things (IoT), Virtual reality has been used for humans and computers naturally and intelligently. A study by Lv (2020) shows that the interaction, simulated natural state and 3D environment can be formed at the display terminal through the processing and operation of information by a computer program, which can make people feel immersed which means that emerging VR to IoT could be very beneficial to promote industrial applications and enhance human activities.

Portman, Natapov and Fisher-Gewirtzman (2015) reviewed the VR Environments for research and teaching for different disciplines such as architecture and environmental planning. For architecture, VR contributes to the wide development for a practitioner to use VR in their work while for Environmental planning VR is less frequently used since an exception to climate change issues is one of the contributing factors to consider. VR was used that creates a framework for testing the overall validity of the proposed plan whether from an architectural design and environmental planning.

These VR developments give the researchers some ideas as an inspiration on how to develop an effective and efficient project using the technology.

## METHODOLOGY

*Software Development*

The study uses a Modified Iterative Development as a guide for the design and development of the VR 3D Model and its environment. Iterative development is a procedure on which there is a breaking down of the computer program advancement of a huge application into little portions. In the iterative development process, the highlighted code is planned, created, and tried in rehashed cycles. In each repetition, there is re-designing of additional features, developed and verified until it is ready to be deployed or installed. Since this project used Modified Iterative Development, there are some changes of the iteration on the stages to ensure that the program follows the desired outcomes set by the developers. Figure 1 illustrates the Modified Iterative Development to produce an increment project that is being developed. The process starts with the Planning, Data Gathering and Data Analysis, System Requirements, Designing, Testing, and Evaluation.

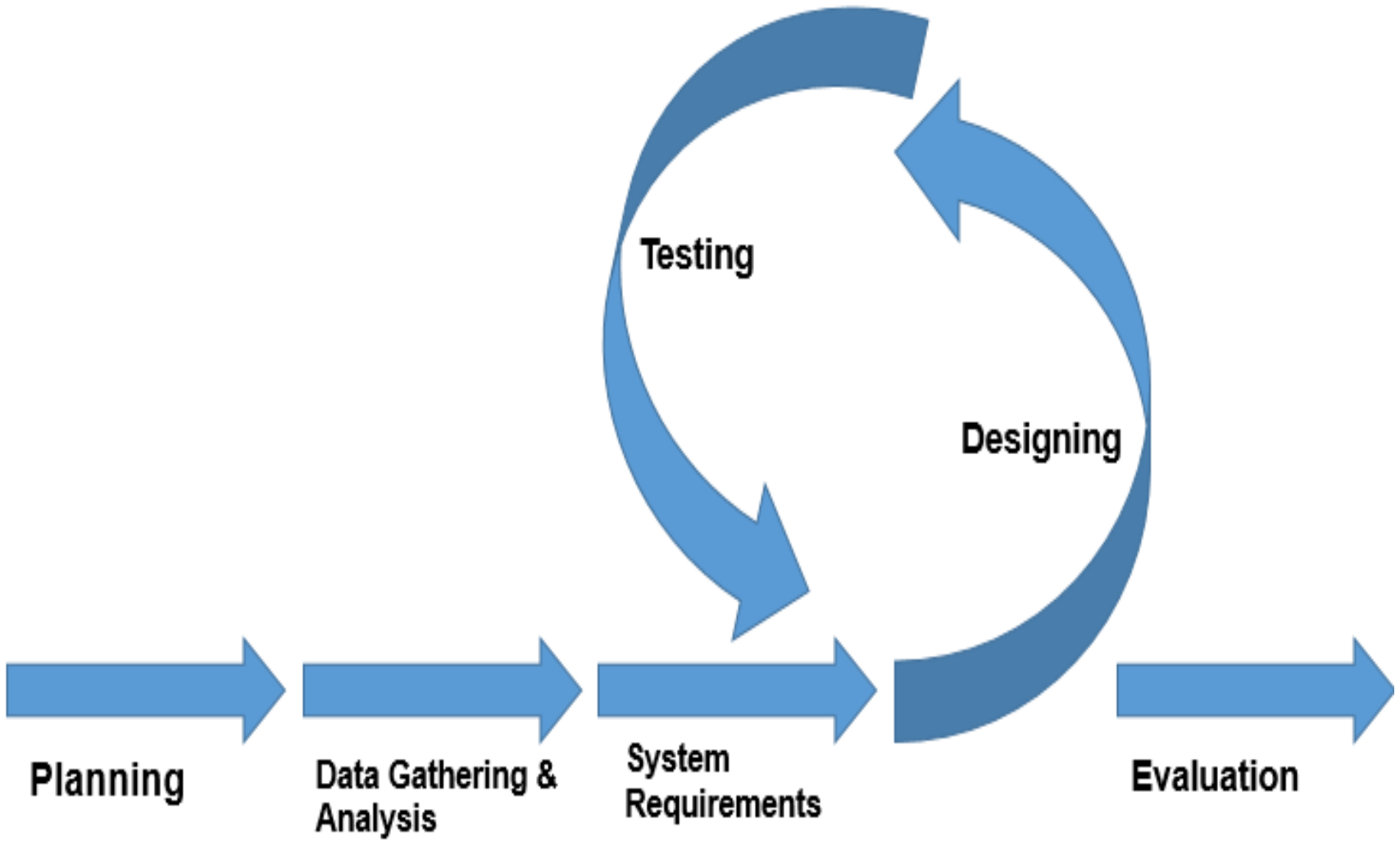

*Figure 1*. Modified Iterative Development

*Planning*

Project planning may be technical procedures in doing projects where required documents are made to guarantee effective extended completion of the project. Documentation incorporates all activities required to characterize, get ready, coordinate, and facilitate extra plans. The project plan clearly characterizes how the project is performed, evaluated, and deployed. In creating the project, thorough planning constitutes primarily the entire actions because this is the process of organizing all the tasks; such that who will do the task and when will be the tasks for the project should be done will be determined. In this part also, the materials and tools to be used for the project are identified. The researchers used the Gantt chart as their guide and basis to create their system. At first, the researchers planned to create a system with the use of Augmented Reality (AR) technology but AR is really expensive and not good for simulating multiple near accurately scaled buildings, that's why they decided to use Virtual Reality instead. After their planning, they did the brainstorming for them to contribute ideas and to give them insights into what are the possible works to do. The researchers had different assignments and roles.

The researchers used the Gantt chart as shown in figure 1 below to determine what work should be done in a specific time frame. Gantt chart really helped the researchers plan and schedule the project effectively. This helped them evaluate the entire duration of the project whether they hit the target or not, this also helped them in managing the dependencies between their assignment and tasks.

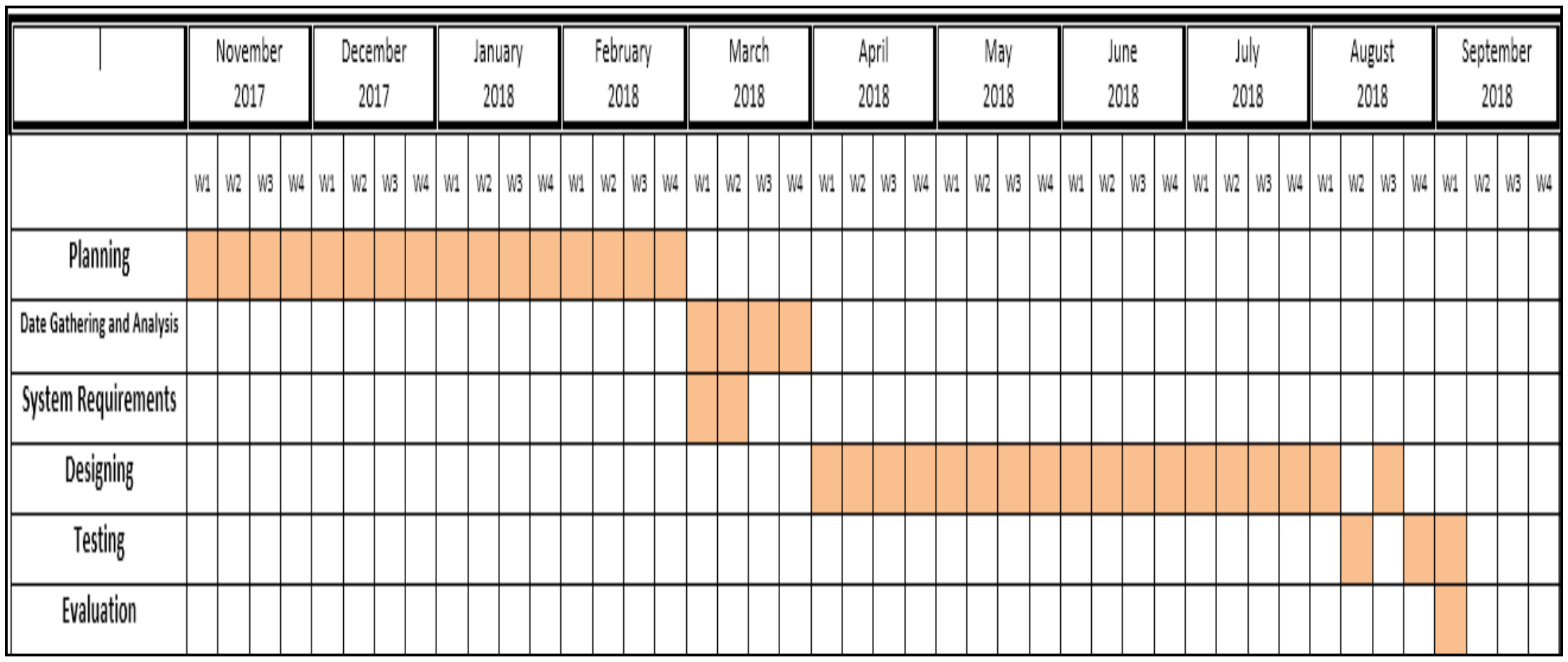

*Figure 2*. GANTT Chart for the Project Development

*Data Gathering and Data Analysis*

The gathering of information is the next step which is vital in interpreting or analyzing structure responses that increase the core knowledge on the project to develop. The data can be gathered in a diverse way, for example, interviews, observations, gathering of receipts and forms, surveys, experiments, and other related activities. The data or information that has been gathered must be summarized, interpreted, and analyzed before having conclusions (UniSA, 2016).

In gathering data, the researchers used the Survey Method using questionnaires to be answered by the students. The questionnaire consists of two questions that are related to the study. The overall response of the students is used as the basis of the project to pursue allowing the project to gather relevant data to determine its validity and success. Data analysis is the next step after data gathering. In this part, the project analyzed all the data that were gathered, meaning the answers from the survey were organized and then interpreted. The results then served as a guide for the validity and pursuit of the project.

The researchers conducted a survey of at least 10 percent of the total student population of each department to determine if the miniature was enough for viewing the inside and outside structure of the building. Out of 107 total respondents of each department, 72 students said that the architectural miniature is not enough to visualize the present structure of the school, and 79 agreed that there should be an alternative way to visualize the existing structure of Northern Bukidnon Community College. Majority of the student favors about the idea of having a new alternative way in visualizing the present structure of the school.

*System Requirements*

System requirements refer to the equipment and program components of a computer framework that are required to introduce and utilize a computer program effectively. This is used to know whether the computer components installed would suffice the necessary requirements for the software to properly run on the system (Beal, 2018). Through this procedure, the project identifies and decides for the appropriate qualifications of the software and hardware to be used, because the project requires specific system qualifications to be followed. The researchers used these software applications in creating their system, these are Unity as their Engine, Monodevelop-Unity as their Text editor, C# as their Programming Language, and Kitkat for the Android OS.

*Designing*

Designing software is a long process since it requires process defining of the things to use that includes its functions, the objects to consider, the user interface, the general structures which are based on the user requirements (UCAR, 2018). Designing is the most interesting part of all the processes because the project is VR architectural simulation which involves the creation and designing of buildings including coding. The researchers used TinkerCAD to create and model the buildings and developed in the Unity software and then converted the assets from OBJ into FBX file with the use of Autodesk FBX Converter 2013.

*Testing*

Testing is the method or strategy of finding mistakes or possible errors in a program application so that it concurs with the user's necessity (Coleman, 2017). Testing follows right after design. This is the process where to measure the usability and design of the system. The researchers went to the surrounding High School Institutions of NBCC to test their project by the Grade 12 senior high students as the prospective college freshmen if it's functional and if there's an existing error during the testing period. They provided VR gear or goggles and a supported Android phone to be used to test the VR App.

*Evaluation*

Evaluation is very important to software development since it examines whether a program follows its purpose whether it reaches the desired outcome and to ensure sustainability (Patton, 1987). The evaluation part gives the researchers an idea of whether the developed application is ready to be used and to know moreover in the event that it meets its reason. The researchers evaluate the project through user experience which allows the Grade 12 senior high school students to try and evaluate the system through writing comments and feedback about the system. Table 1 shows the questions of the System Usability Scale (SUS) for evaluation.

Table 1. System Usability Scale for Evaluation

| Questions | 1 | 2 | 3 | 4 | 5 |
|---|---|---|---|---|---|
| 1. I think that I would like to use this application frequently | | | | | |
| 2. I found the application very beneficial | | | | | |
| 3. I thought the application was easy to use | | | | | |
| 4. I think that I would not need the support of a technical person to be able to use this application. | | | | | |
| 5. I found the various functions in this application were well united. | | | | | |
| 6. I thought I can work with fewer mistakes with this application. | | | | | |
| 7. I would imagine that most people would learn to use this app very quickly | | | | | |
| 8. I found the application very useful | | | | | |
| 9. I felt very confident in using the application. | | | | | |
| 10. I don't need to learn a lot of things before I could get going with this application. | | | | | |

## RESULTS AND DISCUSSION

The researchers took some pictures of the buildings as a basis for their design. Figure 3 shows some of the pictures of the buildings of the college taken by the researchers.

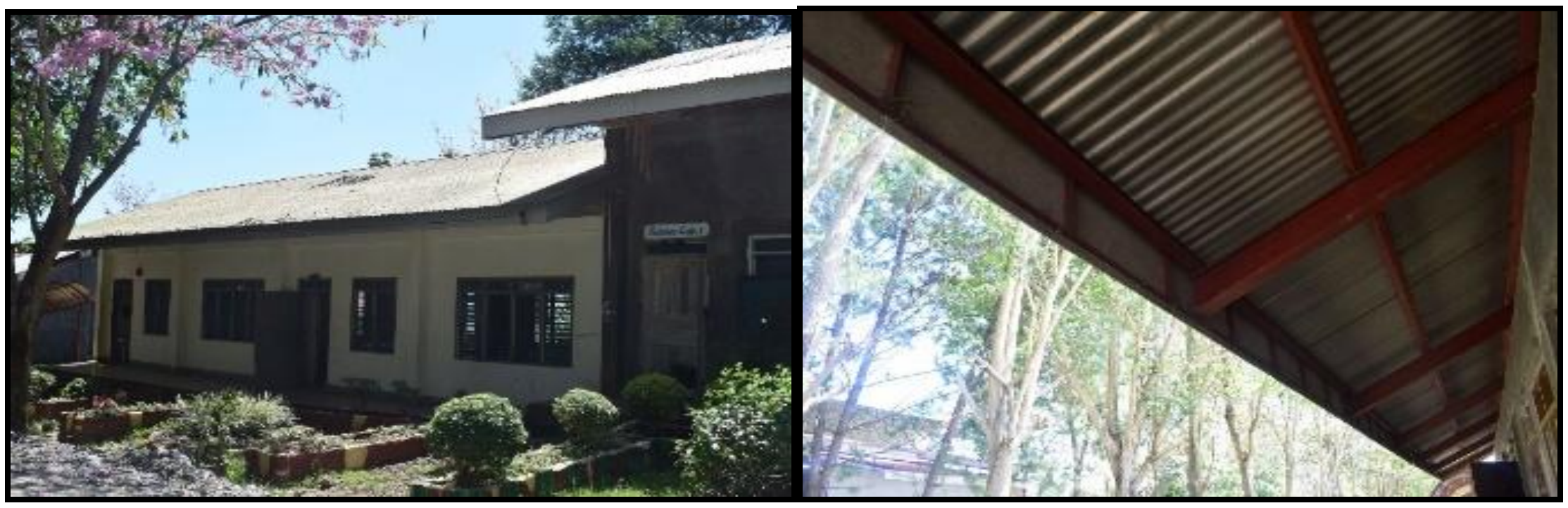

*Figure 3*. Sample Building and Inner Structures

*Software Requirements*

A software requirement is a document that describes the intended purpose requirements and nature of the product to be developed.

Table 2. The Software Requirements for the Development of Virtual Reality Architectural Simulation

| Software | Specifications |
|---|---|
| Game Engine | Unity 5.6.6p3 |
| Script Editor | MonoDevelop |
| Programming Language | C# |
| 3D Model Type Converter | Autodesk FBX Converter 2013 |
| Web Browser | Google Chrome 10/Firefox 4 |
| 3D Modelling Tool | TinkerCAD |
| Android SDK | Android Studio 3.1.4 |

Table 2 shows the Software requirement that was used in creating the project. They used TinkerCAD as their 3D Modeling Tool because it is free and an easy browser-based 3D design and modeling tool and also Unity game engine because this supports the Google Cardboard plugin and is compatible with the programming language that they used.

For Script Editor, the researchers used MonoDevelop. C# is the Programming language used because it is one of the supported programming languages in Unity and it is the most commonly used programming language by other programmers. Autodesk FBX Converter 2013 was used by the researchers to convert OBJ (Object) files to FBX (Filmbox) files. By default, TinkerCAD exports 3D models. STL (Standard Triangle Language), Scalable Vector Graphics (.svg), and Object (.obj) type, but the researchers prefer (.fbx) type to be imported to unity because through converting the 3D model type, the materials of the 3d model are joined as one with the mesh. Unlike in OBJ files, the materials are separated from the mesh and are saved as MTL (material) files.

*Design*

The most interesting and challenging part of the whole development even though the actual scale and measurements of the buildings are not accurate, still looks closer to reality. Figure 4 shows the Guardhouse that is made up of woods and the windows are made up of jalousie. This actual structure for this building was functional for the school's guards.

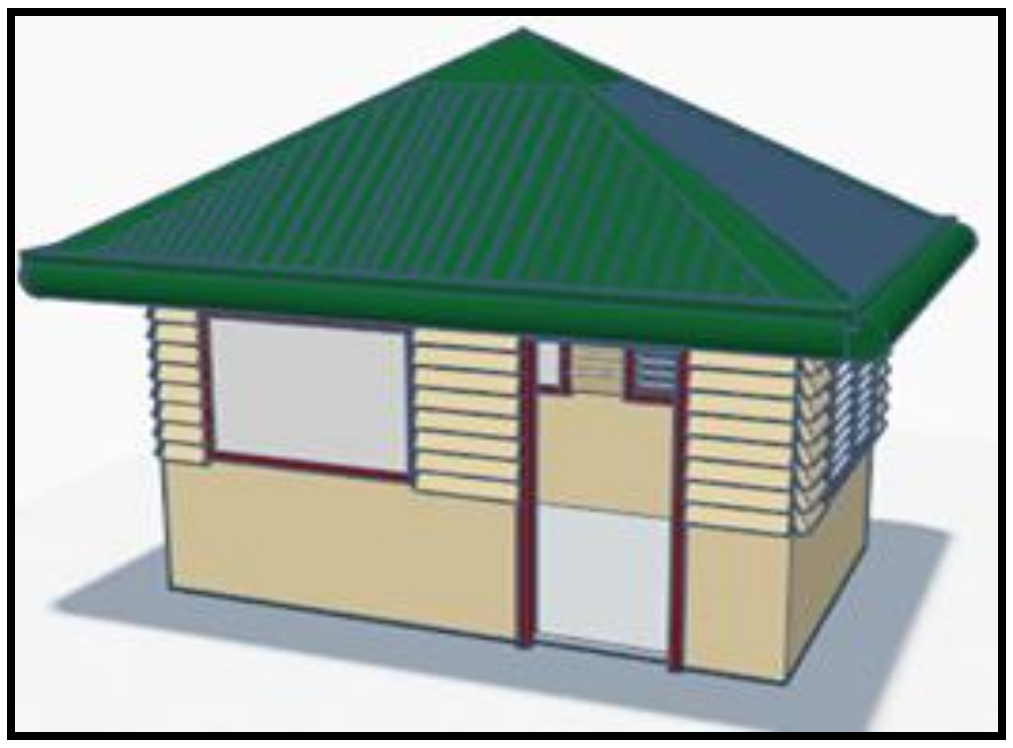

*Figure 4*. 3D Model Guard House

Figure 5 shows the Covered court of the school where most of the school programs and events happen. This was patterned from the actually covered court of the college.

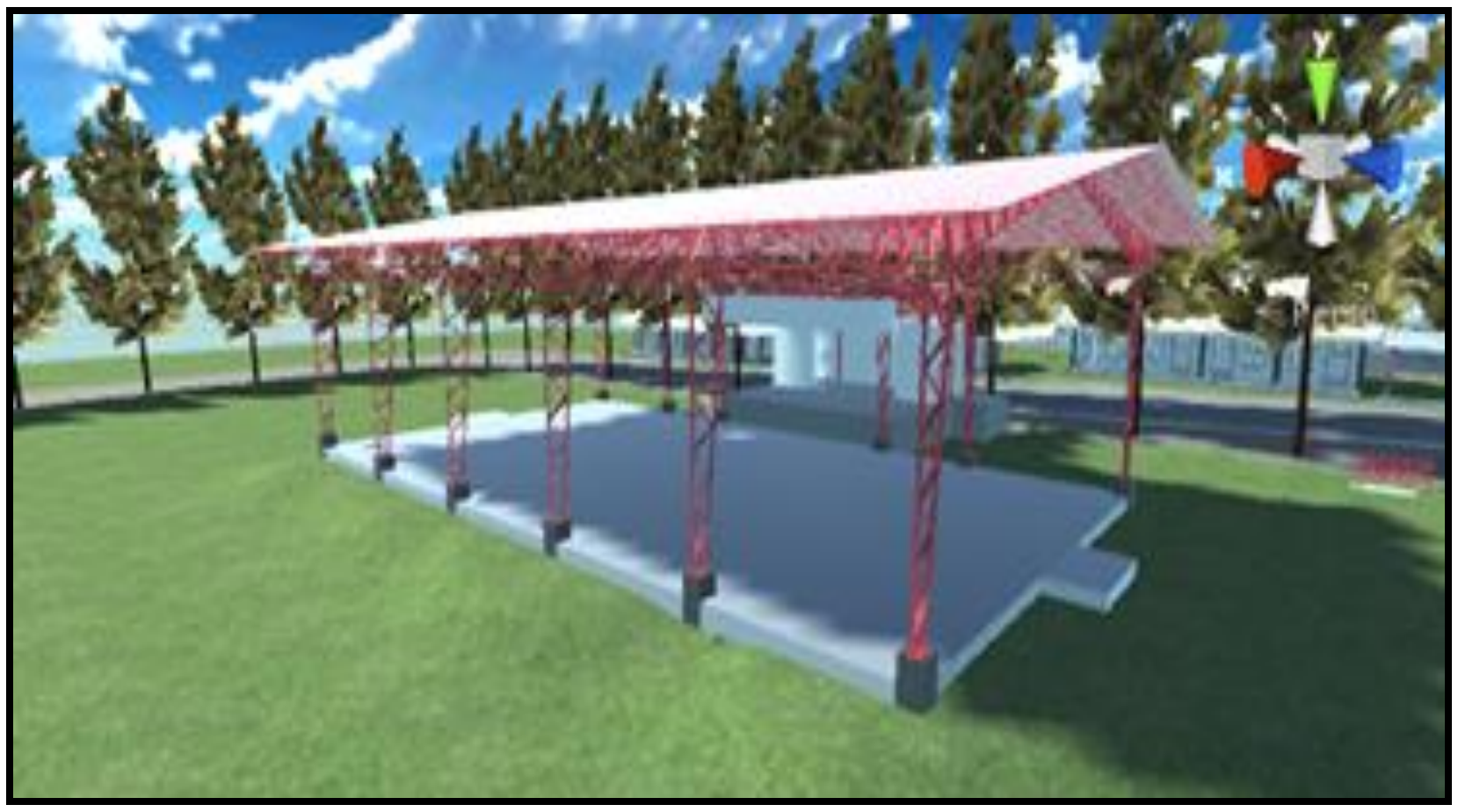

*Figure 5*. 3D Model Covered Court

Figure 6 shows the sample lecture rooms. These rooms are also concrete just like the previous classrooms and are used for student's classes.

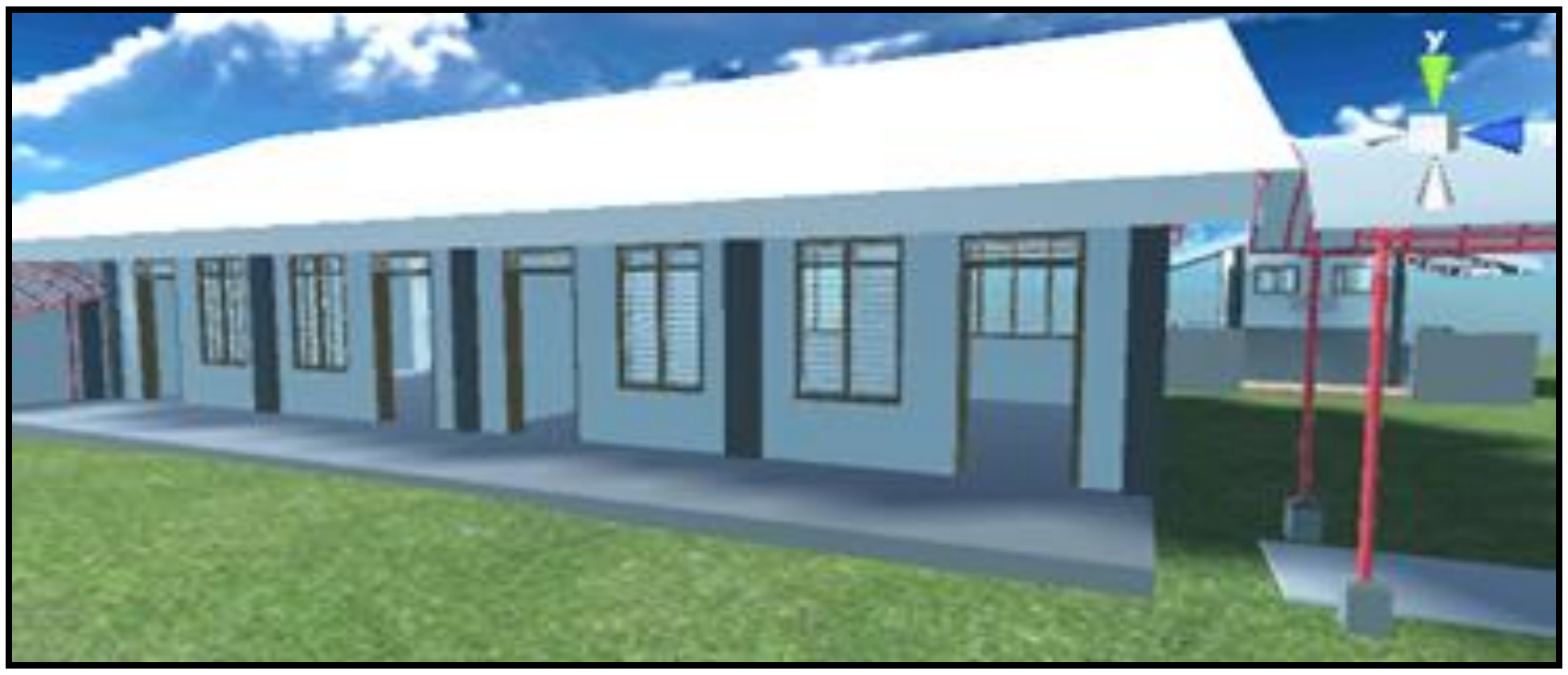

*Figure 6*. 3D Model Classroom

Figure 7 shows the School Library where the students study and conduct their research and assignments. The library is concrete, the windows are made up of jalousie and the doors are made up of metals.

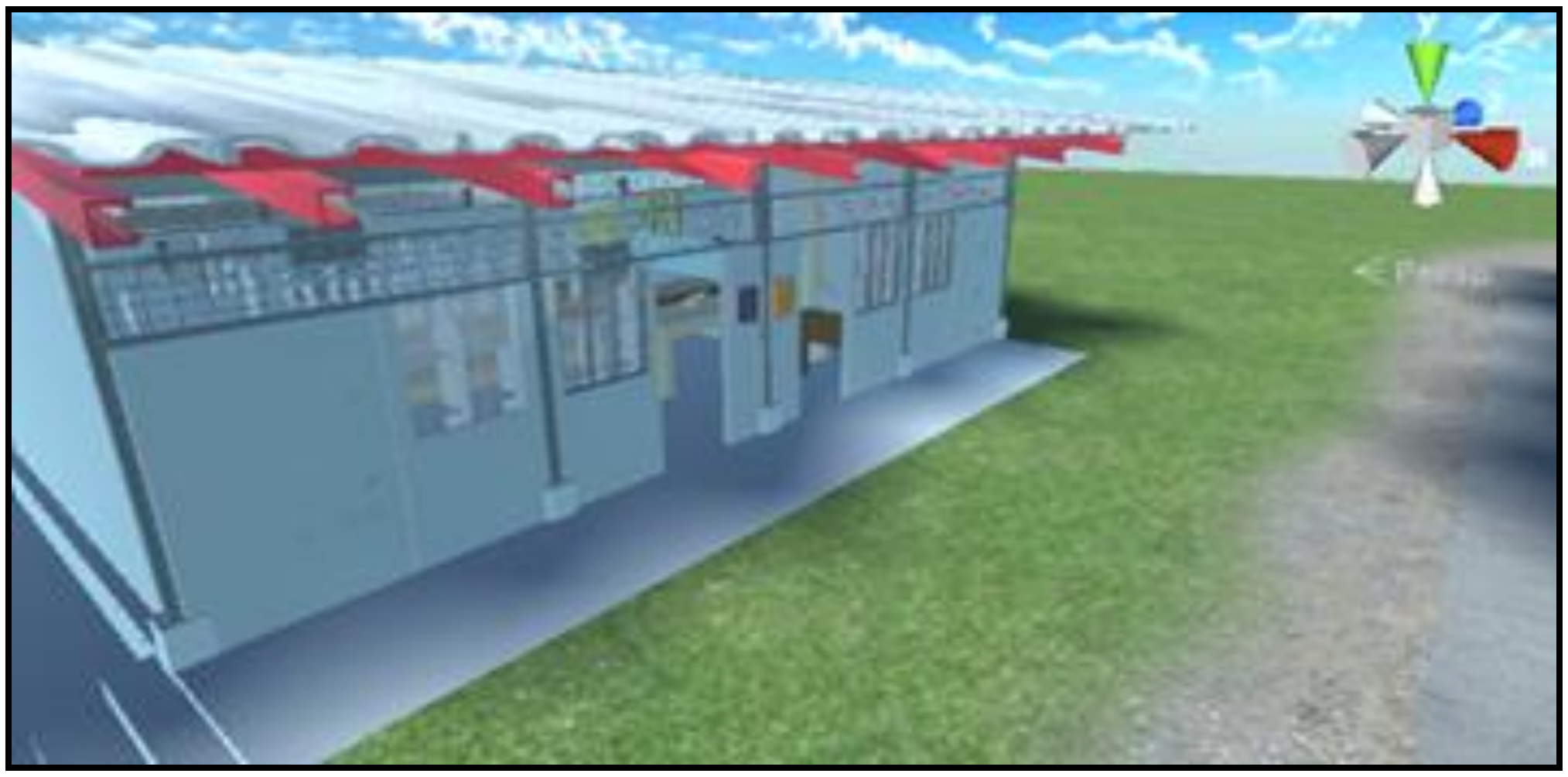

*Figure 7*. 3D Model Library

Figure 8 shows the 3 Story Building, composed of three (3) floors, with five (5) rooms on the second and third floor, while there are six (6) rooms on the first floor, composed of three (3) classrooms, a clinic, SAS office and the Faculty office of the Business Administration Department

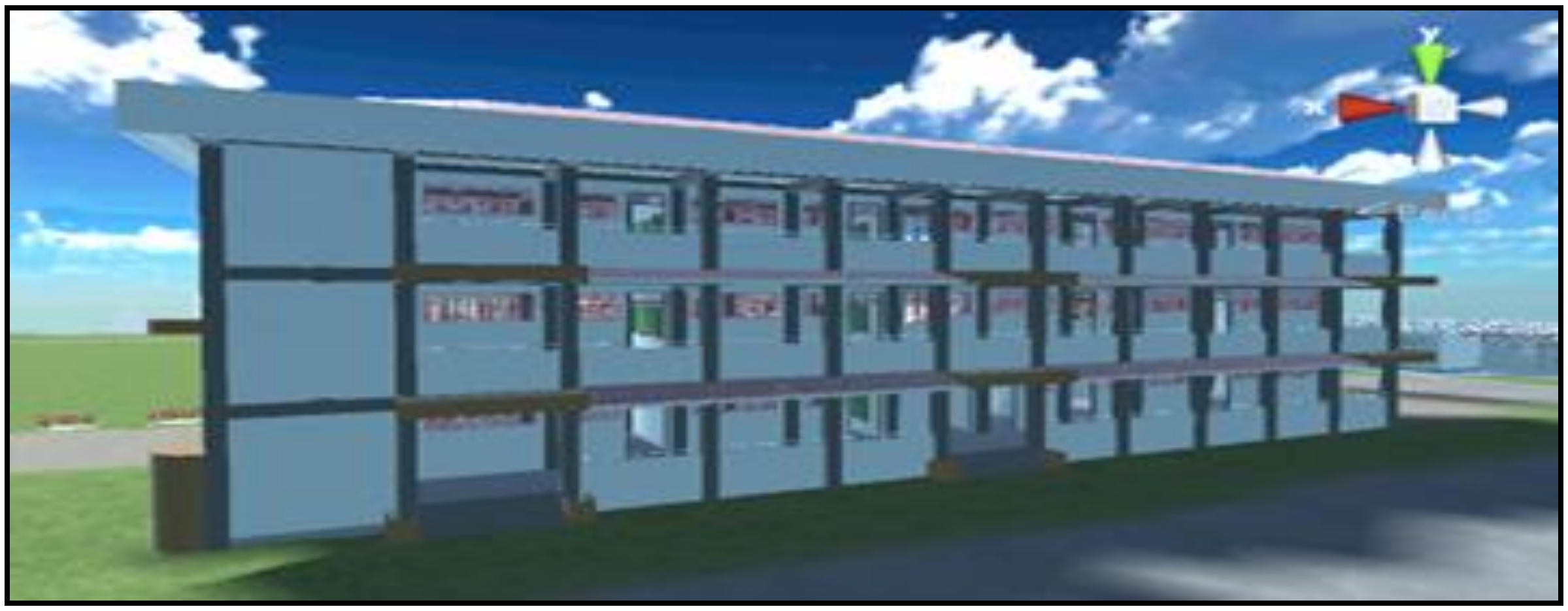

*Figure 8*. 3D Model of a 3 Story Building

Figure 9 shows the Splash Screen of the VR Application while loading and preparing the necessary components for the VR application to work. This flash screen contains the logo that comprises of a VR Glass with letters NBCC representing the college, this is one way of promoting the VR Application to the users while waiting for the necessary components to be loaded.

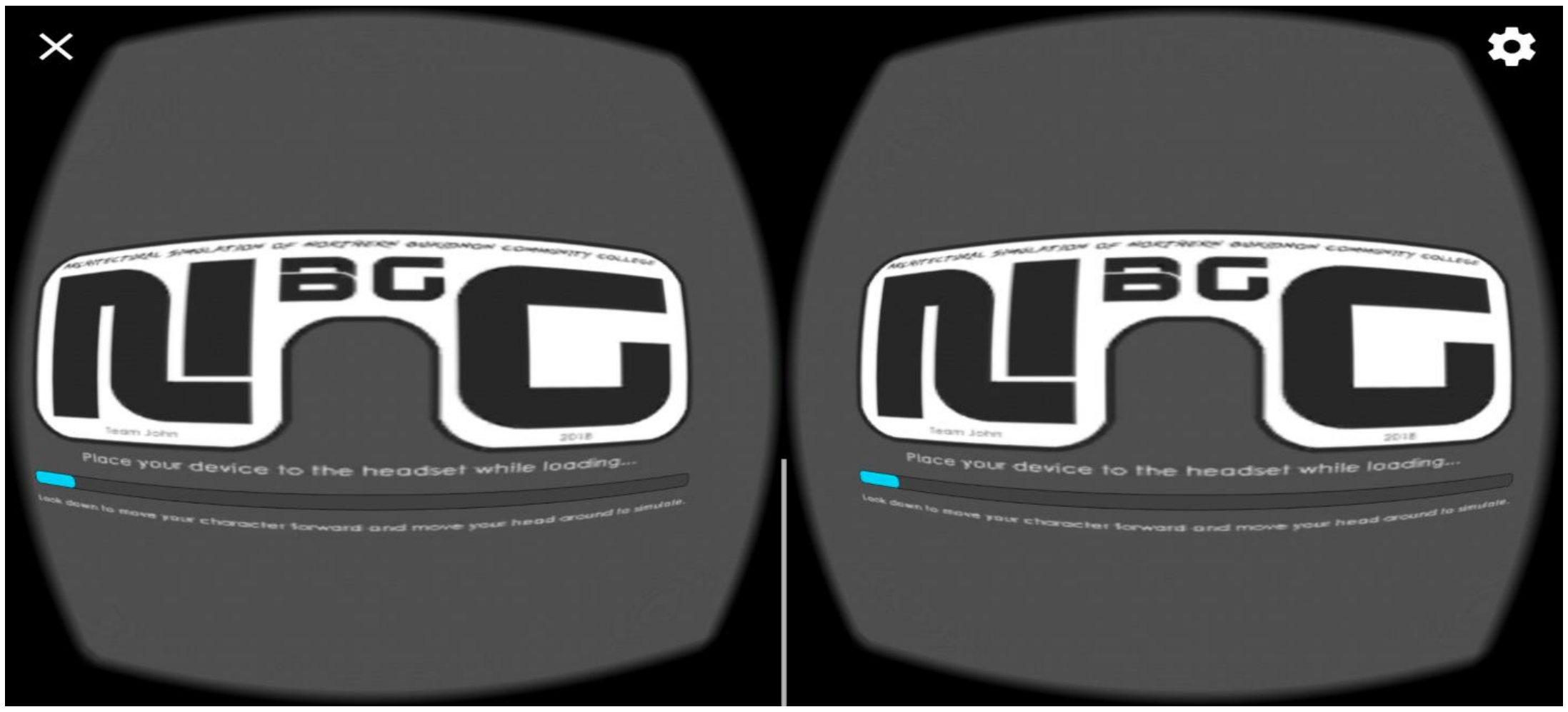

*Figure 9*. Splash Screen

Figure 10 and 11 shows the in-game of the application. The screen is split into two but using the VR Viewer like Google Cardboard, Google Daydream, and VR Headset will merge these two screens into one for a better vision and experience. The user can navigate around by 360 degrees and can move forward by looking 30 degrees below.

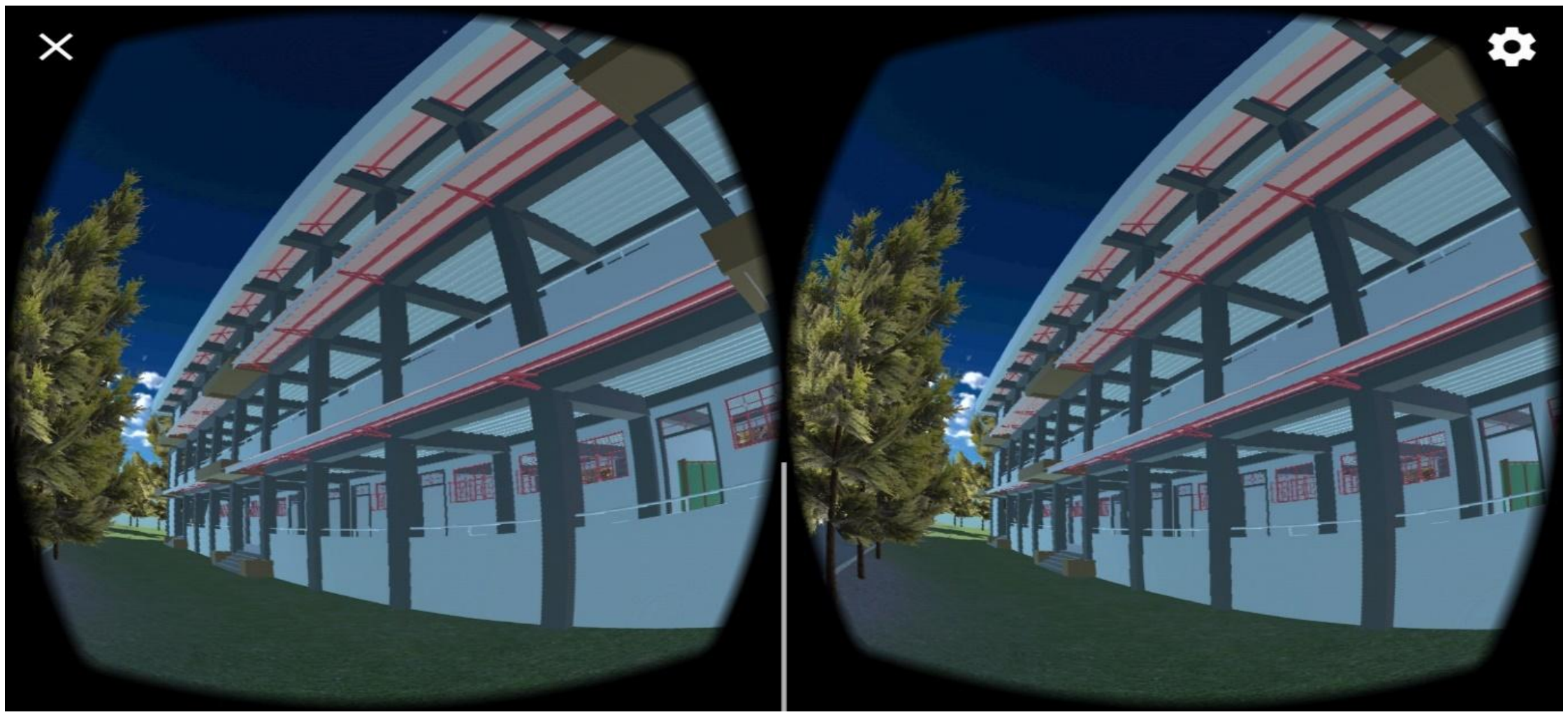

*Figure 10*. In-game 3D Model 3 Story Building

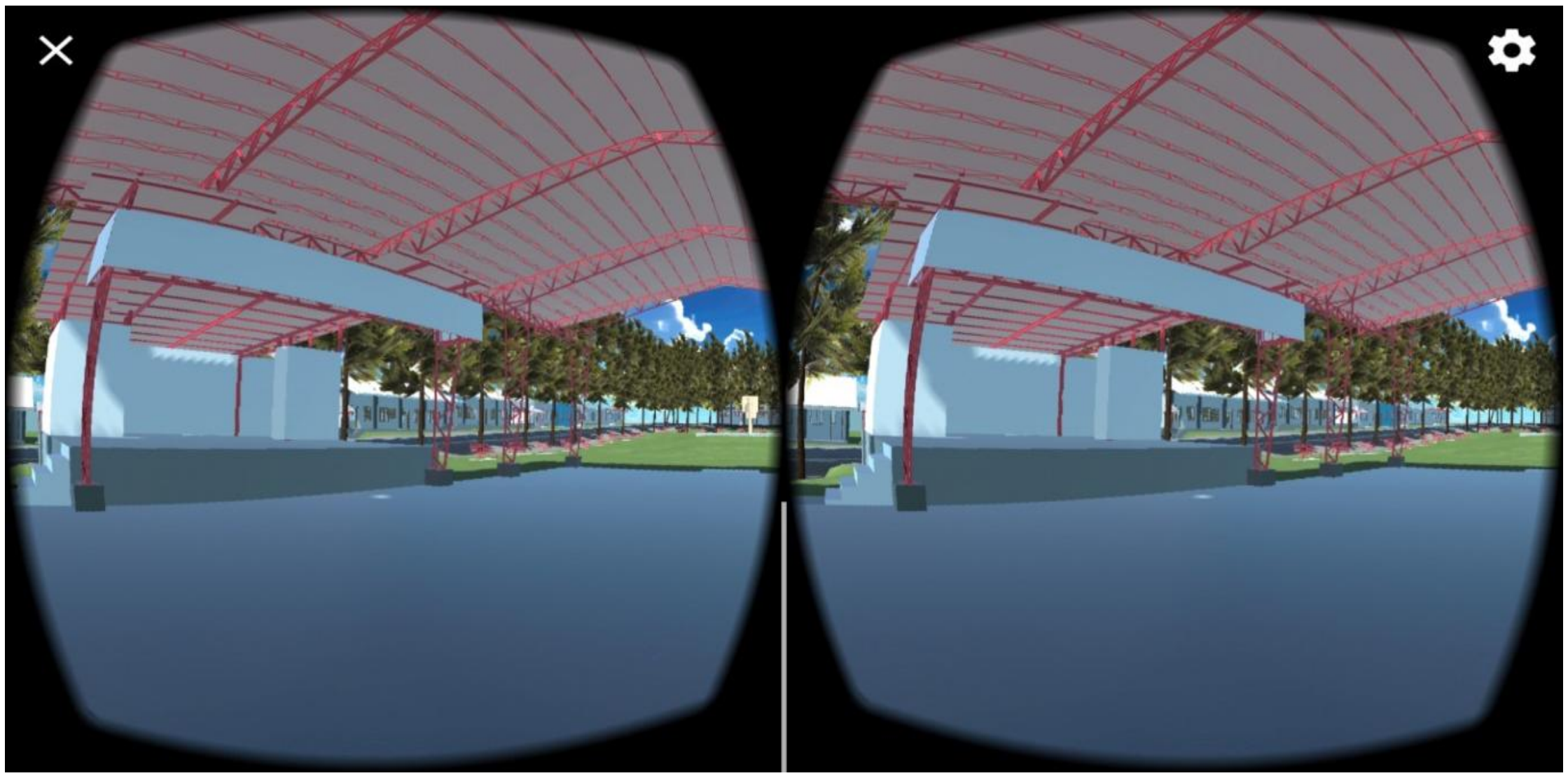

*Figure 11*. In-game 3D Model Covered Court

*Testing*

Testing is the process where the researchers let other users test their projects. Their target respondents are Grade 12 Senior High students as shown in Figure 12 below because these students are the possible future students of the college. The researchers have chosen the four (4) different High School Institutions that surround the college to test their VR application. The researchers sent a letter to each designated principal asking permission to conduct a usability scale test and fortunately all principals in each school gave them permission to do so. Before the testing part, the researchers conducted a short lecture about the project and then allow the students to test the VR Application. Some grade 12 students tried to use the application using VR Googles.

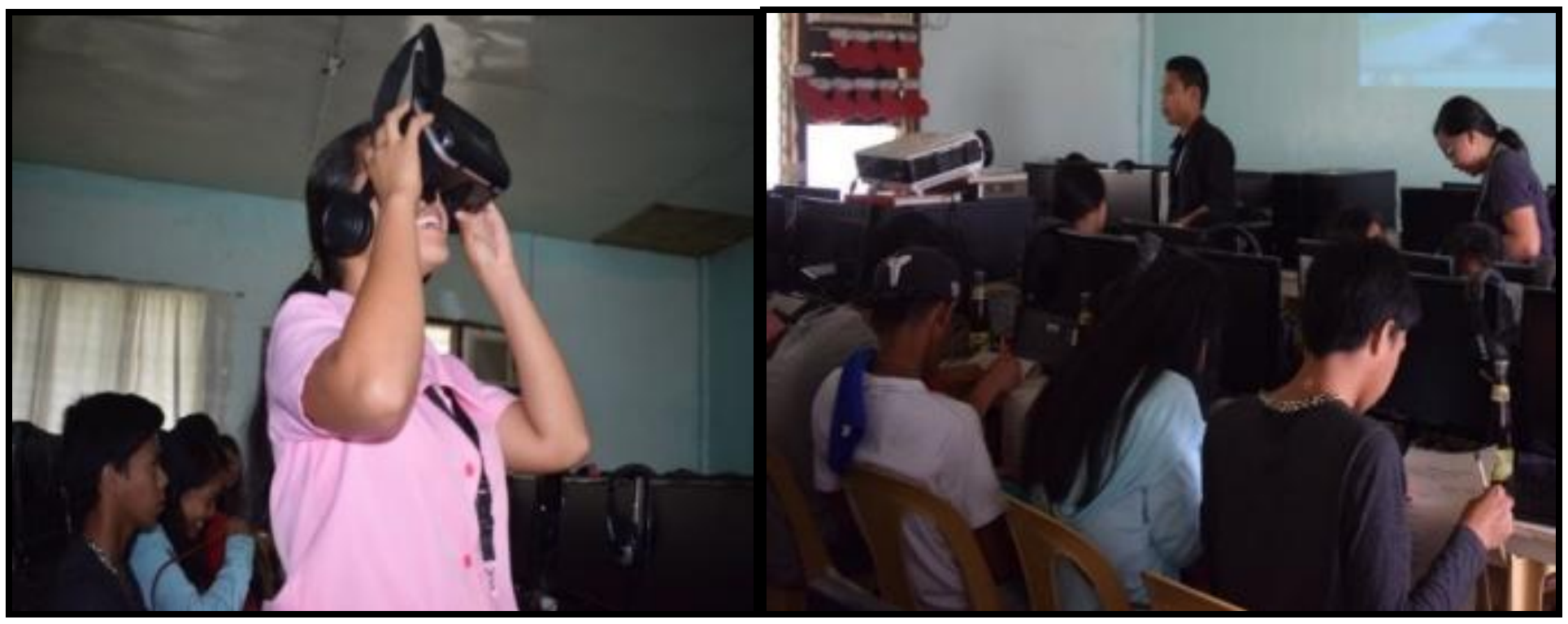

*Figure 12*. Testing at Sankanan National High School

Figure 13 shows the Grade 12 Students of Manolo Fortich National High School testing and evaluating the system seriously. Some of the students enjoyed the experience and invite others to try.

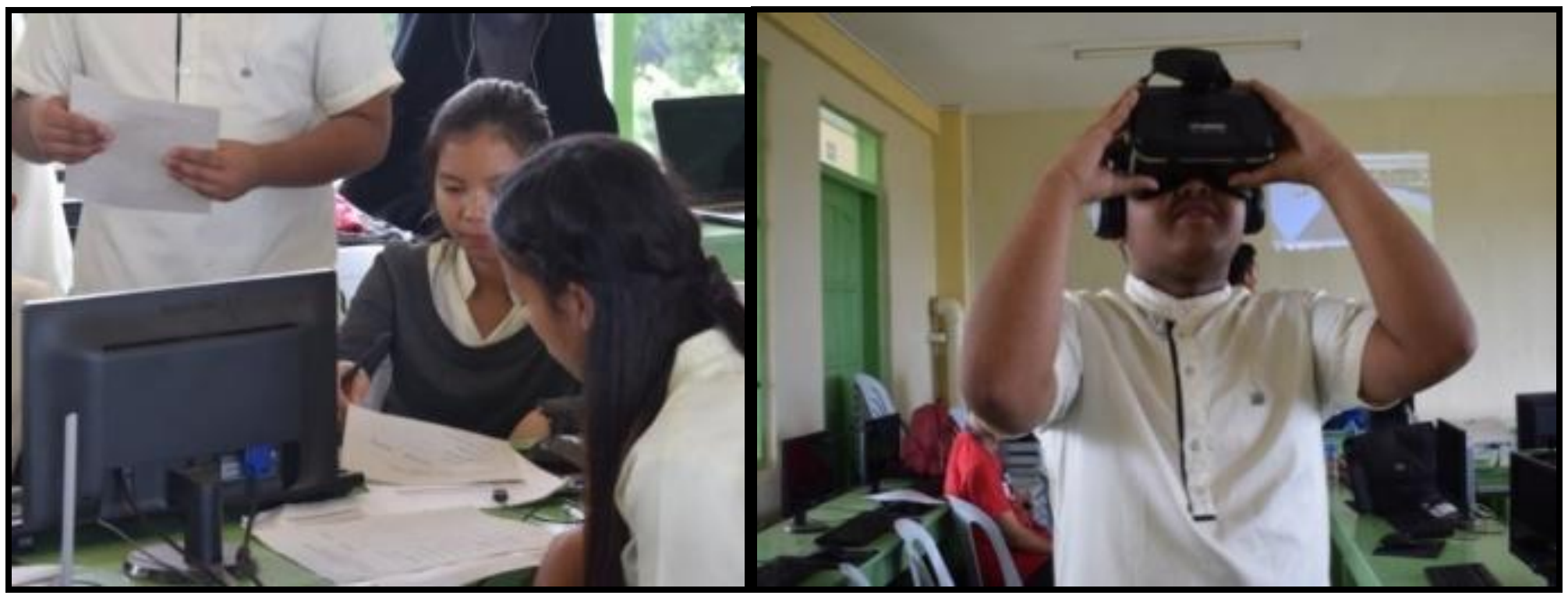

*Figure 13*. Testing at Manolo Fortich National High School

Figure 14 shows the Grade 12 Students of Dalirig National High School testing the system very well and evaluate the system after experiencing the Virtual Reality.

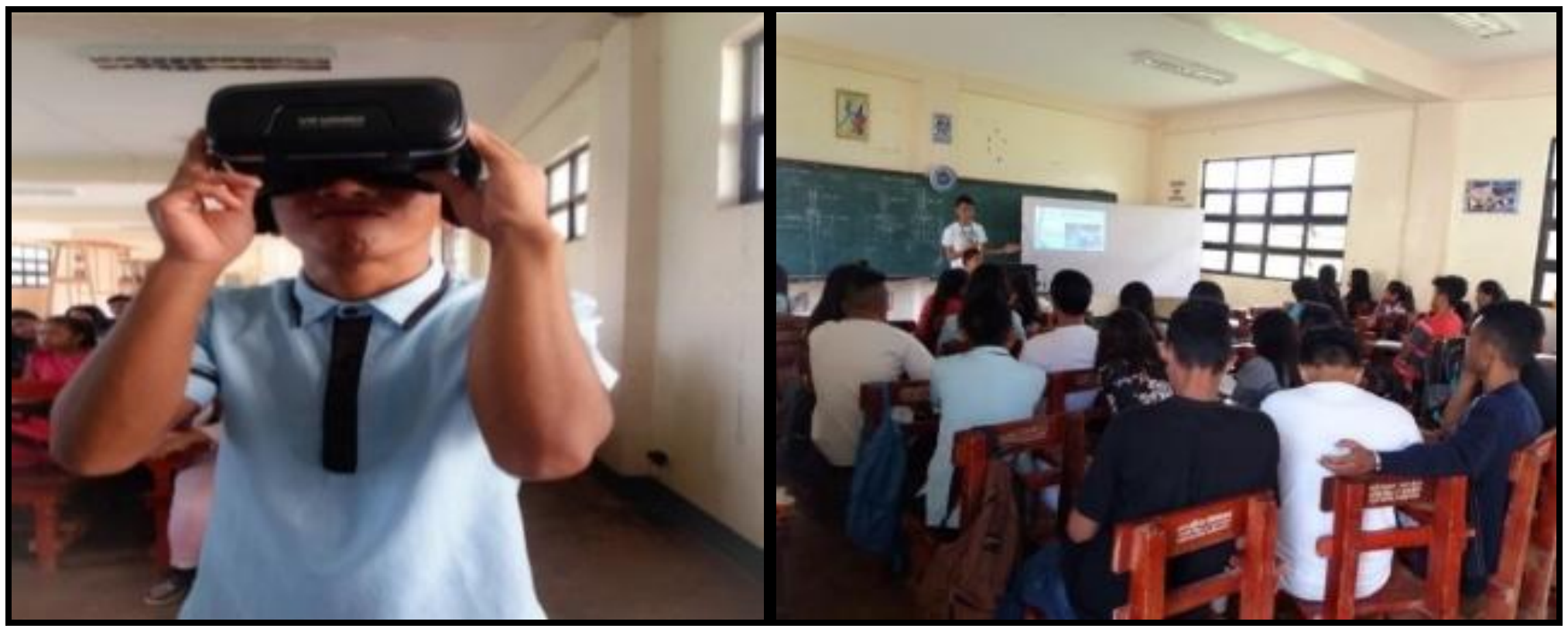

*Figure 14*. Testing at Dalirig National High School

And lastly, figure 15 shows the students of Alae National High School trying the VR Goggles for the first time and enjoyed the experience.

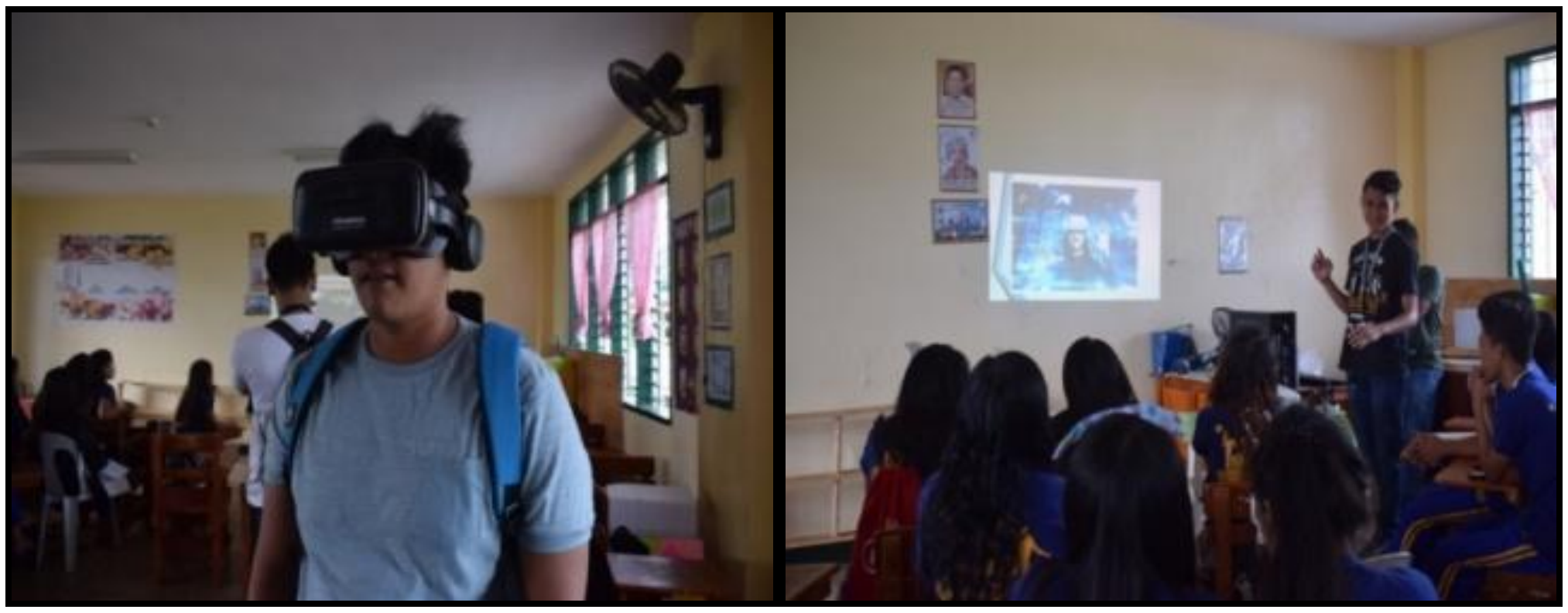

*Figure 15*. Testing at Alae National High School

*Evaluation*

Evaluation is the next process after testing in which the students rate the functionality and at the same the usability of the project (Table 3). Through this, the researchers will be able to determine if their study is relevant and efficient.

Table 3. The Likert Table for the Qualitative Interpretation to be used in interpreting results of the survey.

| Values | Qualitative Interpretation |
|---|---|
| 1.00 – 1.79 | Poor |
| 1.80 – 2.59 | Fair |
| 2.60 – 3.39 | Good |
| 3.40 –4.19 | Very Good |
| 4.20 – 5.00 | Excellent |

The researchers used the qualitative Likert Table of Sorrel Brown (2010) to interpret the result of the evaluation. The range is evenly distributed to ensure that specific results are given consideration.

Table 4. The Overall Evaluation Results

| Sankanan NHS | Manolo Fortich NHS | Dalirig NHS | Alae NHS | Over-all Average |
|---|---|---|---|---|
| 4.70 | 4.50 | 4.40 | 4.40 | 4.50 |

Table 4 as shown above presents the overall VR project rating, based on the results from the four (4) high school institutions where the researchers conducted their testing and evaluation, the VR Application got consistent excellent remarks. Among the four (4) schools, Dalirig and Alae National High Schools rated the VR App least because some students were not given ample time to experience the VR App very well due to the short period of time given to the researchers as permitted by the school principals and the student's excitement were shortened because they have not completed roaming around on all buildings. But somehow the average of 4.4 falls inside the excellent range and it is a great achievement for the researchers.

## CONCLUSIONS AND RECOMMENDATIONS

In this paper, the buildings and structures of NBCC were designed and developed through a Virtual Reality Platform allowing students from different secondary schools that are geographically near to the college to experience the feeling to be in the school without being able to set a step in physically. Using VR Gadgets in navigating buildings is still new from the community which makes the VR application a hit to those who use it. The Application was evaluated personally by every student from different schools and rated a very generous score average of 90% or a point score of 4.5 which means Excellent using a System Usability Scale. With this advancement of technologies, VR plays a vital part in allowing people to see what's inside the building and navigate around it without being physically present in the place.

This study is limited only to simulating the architectural buildings of a Community College in Northern Bukidnon, Philippines, and currently has no controllers used and also, no sound effects. The researchers recommend to future researchers to continue developing a project, to add doors to the rooms, landscapes, mini-map UI, and improve the landform. This is to make the project look closer to the real feature of the college. It is recommended also to add controllers to interact with the objects inside the Virtual Environment. The researchers hope that the future researchers will be courageous enough to continue this project and will not be limited only on viewing the buildings but also adding virtual human to stand as a tour guide when using the VR App.

## IMPLICATIONS

The study enabled viewing and designing of school structures using Virtual Reality technology. With this study, future developments of NBCC in terms of structural design will be easier to visualize, the institution will have convenience on developing its establishment which would allow encouraging more students to enroll. This would contribute to an effective marketing strategy. This study also allows other students to experience the advanced technology of virtual reality, which would interest them to explore more technological ideas and not limit their learning. To Future Researchers, the success and limitation of this study will allow future researchers to scrutinize the process and develop better methods and other systems to improve the field of research of information technology.

## ACKNOWLEDGEMENT

The researchers acknowledge the support of the following people and institutions who have taken part in the success of the project: Kouim Joshua Decena, Vince April Q. Baguio, Michael T. Melecio, John Dexter M. Miro, and Ronald John L. Salvacion. The Manolo Fortich National High School, Alae National High School, Dalirig National High School, and Sankanan National High School. The authors also gratefully acknowledge the use of service and facilities of the Northern Bukidnon Community College for the conduct of this study.